\documentclass[twocolumn,prl,superscriptaddress]{revtex4-1}
\pdfoutput=1

\usepackage{graphicx,bm,amsmath,amssymb,color,soul}
\usepackage{hyperref}
\usepackage{cleveref}

\let\hide\iffalse

\renewcommand{\cite}[1]{{\color{magenta}[#1]}}

\def\bk{{\bf k}}
\def\br{{\bf r}}

\def\bq{{\bf q}}
\def\a{{\alpha}}

\def\bR{{\bf R}}

\def\D{\partial}

\def\w{\omega}

\def\<{\langle}
\def\>{\rangle}
\def\k{\kappa}
\def\ve{\varepsilon}

\def\hH{{\hat{H}}}

\begin{document}

\title{Polarons from first principles, without supercells}

\author{Weng Hong Sio} 
\affiliation{Department of Chemistry, Physical and Theoretical Chemistry,
University of Oxford, South Parks Road, Oxford, OX1 3QZ, UK}
\affiliation{Department of Materials, University of Oxford, Parks Road, Oxford, OX1 3PH, UK} 
\author{Carla Verdi} 
\altaffiliation[Present address: University of Vienna, Faculty of Physics and Center for
Computational Materials Sciences, Sensengasse 8/12, 1090 Vienna, Austria]{}
\author{Samuel Ponc\'e} 
\author{Feliciano Giustino}
\email{feliciano.giustino@materials.ox.ac.uk}
\affiliation{Department of Materials, University of Oxford, Parks Road, Oxford, OX1 3PH, UK}

\date{\today}

\begin{abstract} 
We develop a formalism and a computational method to study polarons in insulators and
semiconductors from first principles. Unlike in standard calculations requiring large supercells,
we solve a secular equation involving phonons and electron-phonon matrix elements from 
density-functional perturbation theory, in a spirit similar to the Bethe-Salpeter equation 
for excitons. We show that our approach describes seamlessly large and small polarons, 
and we illustrate its capability by calculating wavefunctions, formation energies, and spectral 
decomposition of polarons in LiF and Li$_2$O$_2$.
\end{abstract}

\maketitle

Polarons have been attracting unrelenting attention ever since the polaron concept was formulated 
by Landau a century ago~\citep{Landau1933}. For example polarons inspired the search for high-temperature 
superconducting oxides~\citep{Bednorz1986}, they are considered one of the hallmarks of emergent 
behavior in quantum matter~\citep{Verdi2017,Wang2016,Cancellieri2016,Chen2015,Nie2015}, 
and they have been linked to the extraordinary defect-tolerance of the new metal-halide 
perovskites~\citep{Zhu2016}. At a more fundamental level, the quest for a satisfactory quantum-mechanical 
description of polarons stimulated much progress in diverse areas of theoretical physics. For example 
the solution of the Fr\"ohlich polaron problem by Feynman was a landmark in the development of the path 
integral formulation of quantum mechanics~\citep{Feynman1955}, and the Pekar polaron problem~\citep{Pekar1946} 
found applications in general relativity \citep{Bahrami2014} and quantum state reduction~\citep{Penrose2014}.

In the simplest picture, a self-trapped polaron forms when an excess electron or hole deforms a crystal lattice
so as to create a potential well from which it cannot escape. 
Microscopic models of this effect 
have been developed and investigated in many seminal contributions of the last century~\citep{Pekar1946,
Frohlich1950,Lee1953,Feynman1955,Holstein1959}, and subsequently refined to address increasingly more realistic scenarios,
such as multiple electron bands, dispersive phonons, and transport properties~\citep{Lepine1992,Lepine1994,Alexandrov1999,Alexandrov2000,Alexandrov2004,
Perroni2004,Gunnarsson2004,Cruzeiro2000,Alexandrov2003,Hannewald2009}.
More recently, significant progress in the theory of polarons has been 
achieved with the development of numerical many-body techniques,
such as exact diagonalization~\citep{Fehske2007}, renormalization group~\citep{Jeckelmann1998,Grusdt2016}, 
continuous-time quantum Monte Carlo~\citep{Kornilovitch2007},  
and diagrammatic Monte Carlo methods~\citep{Mishchenko2000,Prokofev2008,Cesare2018}.
Comprehensive reviews of the field can be found in Refs.~\onlinecite{Alexandrov2007,Devreese2009,Alexandrov2010,Emin2013,Devreese2016}.

The common denominator to most theoretical studies on polarons is that they focus on idealized
mathematical models, for example the Fr\"ohlich Hamiltonian~\citep{Frohlich1950} and the Holstein 
Hamiltonian~\citep{Holstein1959}, which describe a free or a tightly-bound electron interacting with 
a dispersionless optical phonon, respectively. These models offer an ideal testbed for 
methodological development, and shaped our current understanding of polarons. However, they are not 
suitable for studying {\it real} materials, as they lack essential features such as complex unit 
cells, band structures, phonon dispersion relations, and realistic electron-phonon coupling matrix 
elements. Furthermore, such models are not transferable to complex systems such as surfaces, 
interfaces, low-dimensional materials and heterostructures. Therefore, there
is a need for supplementing correlated methods for polarons with more realistic materials 
parameters, as emphasized by authoritative reviews~\citep{Devreese2009}. 

At the other end of the spectrum, {\it ab initio} calculations based on density functional theory 
(DFT) are ideally positioned to address the complexity of real materials. Indeed, studies of polarons 
under realistic conditions have begun to emerge during the past decade~\citep{Franchini2009,Setvin2014,
Himmetoglu2014,Bondarenko2015,Franchini2017,Franchini2018,Galli2016,kokott2018}. However, also DFT 
faces important limitations: the calculations necessitate large supercells~\citep{Spreafico2014,
Erhart2014,kokott2018}, hence they are prohibitive for intermediate and large polarons which require several 
thousand atoms; local exchange-correlation functionals tend to suppress polaron self-trapping; calculations 
using Hubbard-corrected or hybrid functionals suffer from the sensitivity to the Hubbard parameter 
or the exchange fraction~\citep{kokott2018}. More fundamentally, the relation between DFT calculations 
of polarons and the vast literature on model Hamiltonians remains unclear.

In the present work we wish to overcome these limitations by filling the gap between model Hamiltonians 
and atomistic calculations of polarons. To this aim, we reformulate the direct calculation of polarons
with DFT into a nonlinear eigenvalue problem. The ingredients of this formalism are the band structures, 
phonons, and electron-phonon matrix elements calculated in the crystal unit cell from density functional 
perturbation theory. The solution of this eigenvalue problem yields the formation energy of the polaron, its 
excitation energy, the electronic wavefunctions and atomic displacements, as well as the spectral 
decomposition of the polaron in terms of the underlying electron-phonon coupling mechanisms. We validate 
this methodology by studying two limiting cases, the large electron polaron in LiF and the small electron 
polaron in Li$_2$O$_2$. Complete derivations and extensive benchmarks are reported in the companion 
manuscript, Ref.~\onlinecite{Sio2018}.

We start by considering the DFT total energy of an excess electron added to a crystal with a finite band 
gap. The same reasoning holds for holes~\citep{Sio2018}. The ground state is spin-unpolarized, and we make the working 
assumption that the perturbation to the valence Kohn-Sham (KS) states due to the extra electron can be 
neglected (to be validated {\it a posteriori}). By expanding the total energy in powers of the atomic 
displacements from their equilibrium positions in the ground state, at the lowest order which admits 
non-trivial solutions we obtain:
  \begin{equation}\label{eq.1}
  E = \int \!\! d\br\, \psi^* \hH_{\rm KS}^0 \, \psi \,+ \!\int \!\! d\br\, \frac{\D V_{\rm KS}^0 }
  {\D \tau_s}\, |\psi|^2 \,\Delta \tau_s + \frac{1}{2} C^0_{ss'} \Delta\tau_s \Delta\tau_{s'}.
  \end{equation}
Here the energy $E$ is referred to the ground state, $\psi(\br)$ is the wavefunction of the excess
electron and $\Delta\tau_s$ are the atomic displacements; $s = (\k\a p)$ is a composite index denoting 
the Cartesian coordinate $\a$ of atom $\k$ in the unit cell $p$, and the Einstein summation convention 
is implied. The integrals are over a suitably large Born-von K\'arman supercell. $\hH_{\rm KS}^0$, 
$\D V_{\rm KS}^0/\D \tau_s$, and $C^0_{ss'}$ represent the KS Hamiltonian, the variation of the KS 
potential resulting from an atomic displacement, and the matrix of interatomic force constants, 
respectively. The superscript `0' indicates that these quantities are evaluated in the ground state, 
without excess electron. In Eq.~(\ref{eq.1}) the spurious Hartree and exchange-correlation interactions 
of the polaron with itself and its periodic images are carefully eliminated via a self-interaction 
correction, as discussed in Ref.~\onlinecite{Sio2018}.

The total energy $E$ in Eq.~(\ref{eq.1}) can be regarded as a functional of $\psi$ and $\Delta\tau_s$.
By minimizing this functional subject to the constraint that $\psi$ be normalized, we obtain the
nonlinear eigenvalue problem:
  \begin{eqnarray}\label{eq.2}
  && \hH_{\rm KS}^0\, \psi + \frac{\D V_{\rm KS}^0 }{\D \tau_s} \,\psi \, \Delta \tau_s = \ve \, \psi,\\
  \label{eq.3}
  && \Delta\tau_s = -(C^0)^{-1}_{ss'} \!\int \!d\br\, \frac{\D V_{\rm KS}^0 }{\D \tau_{s'}}\,|\psi|^2,
  \end{eqnarray}
where $\ve$ is the polaron eigenvalue. In principle these equations could be solved in real space,
but in most practical applications this is prohibitive, since the supercell must be large enough 
to accommodate the wavefunction $\psi$. To overcome this obstacle we proceed as in the calculation 
of excitons via the Bethe-Salpeter equation~\citep{Andreoni1995, Rohlfing1998}, that is we expand the 
solution in terms of unperturbed KS states and phonons eigenmodes. To this aim we define $\psi = 
N_p^{-1/2}\sum_{n\bk} A_{n\bk} \,\psi_{n\bk}$ and  $\Delta\tau_s = - 2N_p^{-1}\sum_{\bq\nu} B^*_{\bq\nu}\, 
(\hbar/2M_\k \w_{\bq\nu})^{1/2} e_{\k\a,\bq\nu} \, e^{i\bq\cdot \bR_p}$. Here $N_p$ is the number of unit 
cells in the supercell, $\psi_{n\bk}$ is an unoccupied eigenstate of $\hH_{\rm KS}^0$ for the band $n$ and 
wavevector~$\bk$ with energy $\ve_{n\bk}$, and $e_{\k\a,\bq\nu}$ is the vibrational mode with branch 
$\nu$, wavevector $\bq$, and frequency $\w_{\bq\nu}$, obtained by diagonalizing $C^0_{ss'}$. $\bR_p$ is 
a vector of the direct lattice, and $M_\k$ is the mass of atom $\k$. Using these definitions in
Eqs.~(\ref{eq.2})-(\ref{eq.3}) we obtain a nonlinear eigenvalue problem for the generalized Fourier 
amplitudes $A_{n\bk}$ and $B_{\bq\nu}$:

  \begin{eqnarray} 
  \label{eq.4}  
  &&\frac{2}{N_p}\sum_{\bq m\nu}B_{\bq\nu}\,g^*_{mn\nu}(\bk,\bq)\,A_{m\bk+\bq}=(\ve_{n\bk}-\ve)\,A_{n\bk},\\
  \label{eq.5}
  &&B_{\bq\nu}=\frac{1}{N_p}\sum_{mn\bk}A^*_{m\bk+\bq}\,\frac{g_{mn\nu}(\bk,\bq)}{\hbar\w_{\bq\nu}}\,A_{n\bk},
  \end{eqnarray}

where $g_{mn\nu}(\bk,\bq)$ is the electron-phonon matrix element for the scattering between the
electronic states $|n\bk\>$ and $|m\bk+\bq\>$ via the phonon $\bq\nu$~\citep{Giustino2017}.
Equations~(\ref{eq.4})-(\ref{eq.5}) only require the band structures, phonon dispersions, and matrix 
elements calculated in the {\it unit cell}. In this representation the formation energy $\Delta E_f$
of the polaron, that is the total energy of the self-trapped polaron minus the energy of the
undistorted crystal with an extra electron at the conduction band bottom, reads~\citep{Sio2018}:
  \begin{equation}
  \Delta E_f = \!\frac{1}{N_p} \!\sum_{n\bk}|A_{n\bk}|^2  (\ve_{n\bk}-\ve_{\rm CBM}) 
  -\frac{1}{N_p}\!\sum_{\bq\nu} |B_{\bq\nu}|^2\,\hbar\w_{\bq\nu},
  \end{equation}
where the KS eigenvalue is referred to the conduction band minimum. This expression shows that the polaron
formation energy consists of a positive-definite electronic contribution and a negative-definite
vibrational contribution. This spectral representation allows us to identify $|B_{\bq\nu}|^2$
with the number of phonons taking part in the polaron, as discussed in greater detail
in Ref.~\onlinecite{Sio2018}.

We now illustrate this approach using LiF and Li$_2$O$_2$ as case studies. LiF is a simple salt that 
crystallizes in the rocksalt structure. It is a paradigmatic wide gap polar insulator, and is known 
to host large electron polarons~\citep{Iadonisi1984}. Li$_2$O$_2$ is a prototypical cathode for 
lithium-air batteries, and hosts small electron polarons~\citep{Feng2013}. The structure 
consists of two-dimensional LiO$_2$ layers intercalated by Li planes. We perform calculations 
within the PBE generalized gradient approximation to DFT~\citep{Perdew1996}, 
using planewaves and pseudopotentials as implemented in the Quantum Espresso suite~\citep{Giannozzi2017}. 
We employ optimized norm-conserving Vanderbilt pseudopotentials~\citep{Hamann2013} and planewaves kinetic 
energy cutoffs of 150~Ry and 105~Ry for LiF and Li$_2$O$_2$, respectively. We calculate phonons and 
electron-phonon matrix elements within density functional perturbation theory~\citep{Baroni2001,Giustino2017}, 
and we perform Wannier interpolation of all properties using the Wannier90~\citep{Marzari2012,Mostofi2014} 
and EPW~\citep{Giustino2007,Ponce2016} codes. We employ uniform Brillouin-zone grids in 
Eqs.~(\ref{eq.4})-(\ref{eq.5}), and we solve the linear system via a parallel steepest descent algorithm. 
As the initial seed for $A_{n\bk}$ in Eq.~(\ref{eq.5}) we use a simple Gaussian lineshape~\citep{Sio2018}.

  \begin{figure*}[t]
  \includegraphics[width=\textwidth]{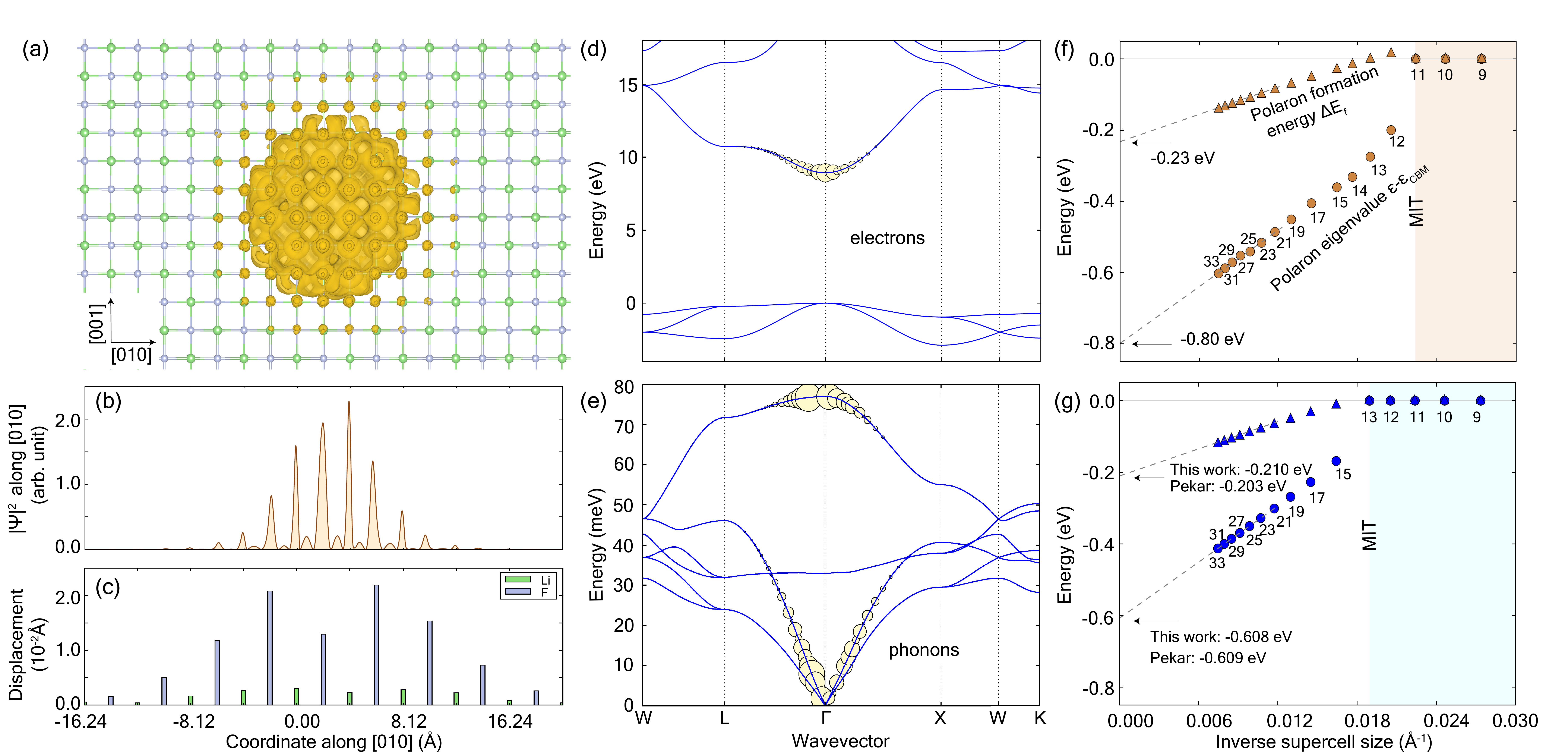}
  \caption{
  \color{black}
  Large electron polaron in LiF. (a) Isosurface of the polaron density $|\psi|^2$ and ball-stick model
  of LiF, with Li in green and F in silver. (b) Cross-section of the polaron
  density $|\psi|^2$ along a [010] line cutting through the center. (c) Modulus of the
  atomic displacements projected along the [010] direction, and taken on a -Li-F- chain of atoms
  nearest to the polaron center. The horizontal axes in (a), (b), and (c) are aligned. (d), (e) Band
  structures and phonon dispersions of LiF, respectively. The Fourier amplitudes $A_{n\bk}$ 
  and $B_{\bq\nu}$ are superimposed to the bands, with the radius of the circle proportional to their
  square modulus. In (d) the zero of the energy is aligned with the valence-band top. 
  (f) Polaron formation energy $\Delta E_f$ and eigenvalue $\varepsilon$ as a function
  of supercell size. The dashed lines are the Makov-Payne extrapolations. The shading indicates
  that no localized solution was found, and MIT stands for metal-to-insulator transition. The numbers
  next to the circles indicate the unit cells in each supercell, e.g. 12 means 12$\times$12$\times$12
  supercell. (g) Polaron energies (triangles) and eigenvalues (circles) obtained 
  with our method using the model Fr\"ohlich electron-phonon coupling compared to the solution 
  of the Pekar polaron model.
  }
  \label{fig1}
  \end{figure*}

Figure~\ref{fig1} shows our results for LiF. In Fig.~1(a) we see that the electron wavefunction
of the polaron extends over $\sim$10~unit cells, therefore we are in the presence of a {\it large} 
polaron. A cross-sectional view in the [010] direction of the same wavefunction is shown in Fig.~1(b). Here we recognize 
an envelope function of approximately Gaussian shape, which modulates the atomic Li and F $2s$
orbitals. From this plot we can quantify the spatial extent of the polaron using the full-width at half maximum, 
$2 r_{\rm p} = 9.0$~\AA. This value is consistent with an earlier semiempirical estimate of 
9.3~\AA~\citep{Iadonisi1984}. In Fig.~1(c) we show the atomic displacements along a line that cuts 
near the center of the polaron. As expected, also the displacements follow an approximately Gaussian 
profile. 

  \begin{figure*}[t]
  \includegraphics[width=\textwidth]{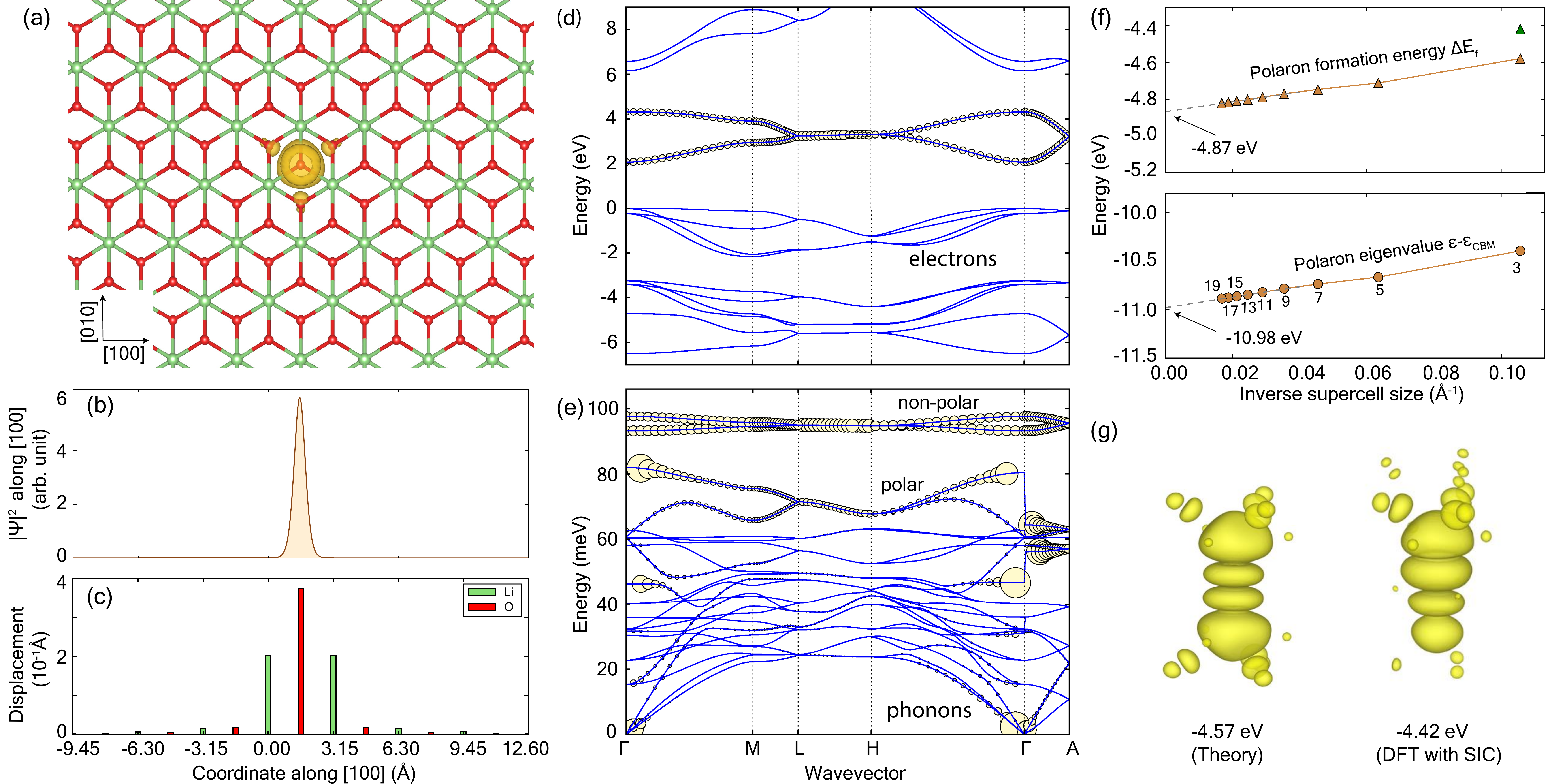}
  \caption{
  Small electron polaron in Li$_2$O$_2$. (a) Isosurface of the polaron density $|\psi|^2$ and model
  of Li$_2$O$_2$, with Li and O atoms in green and red, respectively. (b) Cross-section of
  $|\psi|^2$ along a [100] line cutting through the center. (c) Modulus of the
  atomic displacements along a [001] line passing through the O atom at the center.
  (d), (e) Band structures and phonon dispersion relations of Li$_2$O$_2$, respectively. 
  The amplitudes $A_{n\bk}$ and $B_{\bq\nu}$ are superimposed as circles. The energy zero in (d) 
  is aligned with the valence-band top.  (f) Polaron formation 
  energy and eigenvalue vs.\ supercell size. The dashed gray lines represent the Makov-Payne extrapolations. 
  The green triangle is the result of an explicit DFT calculation from Ref.~\onlinecite{Sio2018}. 
  (g) Comparison between the polaron wavefunction obtained from our method (left), and an explicit 
  DFT calculation (right, Ref.~\onlinecite{Sio2018}), and the corresponding formation energies.
  }
  \label{fig2}
  \end{figure*}

In Fig.~\ref{fig1}(d)-(e) we analyze the composition of the polaron in terms of the Fourier amplitudes
$A_{n\bk}$ and $B_{\bq\nu}$. Panel (d) shows that the electron wavefunction draws weight primarily
from KS states at the bottom of the conduction band. This localization in reciprocal space is consistent
with the highly delocalized nature of the polaron; analogous structures are observed in the related 
physics of Wannier excitons~\citep{Bokdam2016}. In panel (e) we see the phonon eigenmodes
participating in the polaron. There is a strong contribution from the longitudinal optical (LO)
phonons at the zone center, around 77~meV; this is an indication of Fr\"ohlich-type electron-phonon 
coupling. Our approach also reveals a non-negligible contribution from longitudinal acoustic (LA) 
phonons, up to 40~meV. By integrating $|B_{\bq\nu}|^2 $ across the Brillouin zone and summing 
over the phonon branches, we find that the electron polaron in LiF involves $\sim$5 LO phonons and 
$\sim$3 LA phonons, respectively. 

Figure~\ref{fig1}(f) shows the polaron eigenvalue $\ve$ and formation energy $\Delta E_f$
as a function of supercell size. For supercells smaller than 12$\times$12$\times$12 unit
cells the nonlinear eigenproblem in Eqs.~(4)-(5) does not admit localized solutions. This 
can be understood as a manifestation of the Mott transition~\citep{Mott1968}
at a critical density of $4 \cdot 10^{19}$~cm$^{-3}$.
Below this critical density
we find localized solutions of the type shown in Fig.~\ref{fig1}(a), with an energy that scales 
as a constant plus a term proportional to $L^{-1}$, where $L$ is the supercell size. This trend 
is understood as the Madelung energy of a superlattice of polarons. If we extrapolate to 
$L\rightarrow \infty$ we obtain the energy of one isolated polaron~\citep{Madelung1918,Makov1995}. 
In this dilute limit the polaron formation energy is $-0.23$~eV, and the polaron eigenvalue is 
$-0.80$~eV with respect to the conduction band bottom. The ratio between these values follows 
approximately the 1/3 scaling law that is expected for the Pekar polaron, which considers 
exclusively the Fr\"ohlich coupling~\citep{Sio2018,Alexandrov2010}.

To validate our approach for LiF, we performed self-interaction corrected DFT calculations up 
to supercells of size 7$\times$7$\times$7 unit cells, containing up to 686 atoms~\citep{Sio2018}. 
In agreement with the results of Eqs.~(4)-(5), these direct DFT calculations did not yield any localized solutions. 
Calculations for supercells large enough to have localized solutions would be prohibitively expensive, 
as they involve $>3400$~atoms. Therefore, in order to validate our theory in the dilute limit, 
we follow an alternative route and we compare with the prediction of the continuum
Pekar polaron model~\citep{Pekar1946}. To this aim we repeat all calculations in Fig.~\ref{fig1}(f) 
after replacing the band structure by a parabolic model with the DFT effective mass, the phonon 
dispersions by a dispersionless LO mode, and the electron-phonon matrix elements by the long-range 
Fr\"ohlich interaction following Ref.~\onlinecite{Verdi2015}. In Fig.~\ref{fig1}(g) we show that 
our theory reproduces exactly the energetics of the Pekar model in the continuum limit.

Now we move to Li$_2$O$_2$ in Fig.~\ref{fig2}. In this case we find a {\it small} polaron, as
seen from the electron wavefunction in Fig.~\ref{fig2}(a). The polaron is as small as two adjacent
O-$2p$ atomic orbitals [Fig.~\ref{fig2}(g)]; from the cross-sectional view in Fig.~\ref{fig2}(b) we deduce 
a size $2 r_{\rm p} =$~1.3~\AA\ in the (100) plane. Correspondingly the atomic displacements 
are highly localized, and only the first shell of atoms around the polaron center exhibits
non-negligible distortions [Fig.~\ref{fig2}(c)]. The electronic and vibrational Fourier amplitudes 
$A_{n\bk}$ and $B_{\bq\nu}$ of the polarons look very different from the case of LiF in Fig.~\ref{fig1}. 
In fact in Fig.~\ref{fig2}(d) we see that all states of the lowest conduction bands contribute uniformly 
to the polaron wavefunction; similarly phonons from the entire Brillouin zone and from the non-polar and polar branches in Fig.~\ref{fig2}(e) contribute to the atomic displacements. These signatures are 
reminiscent of Holstein-type electron-phonon coupling~\citep{Holstein1959}, albeit with multiple 
electron bands and phonon branches involved. From the square amplitudes 
$|B_{\bq\nu}|^2$ we infer that the small polaron in Li$_2$O$_2$ involves 
$\sim$13~ polar optical phonons centered around 72~meV and $\sim$46~ non-polar optical phonons at energies near 96~meV.

In Fig.~\ref{fig2}(f) we show the polaron eigenvalue and formation energy as a function of supercell
size. In this case we observe the formation of polarons already for 3$\times$3$\times$3 supercells,
and the corresponding critical density for the Mott transition is in the range 
$\sim 10^{21}$~cm$^{-3}$~\citep{Sio2018}. The small electron polaron in Li$_2$O$_2$ is very energetic, 
exhibiting $\Delta E_f =-4.9$~eV and $\varepsilon=-11.0$~eV (with respect to the conduction 
band bottom) in the dilute limit. Our calculated 
formation energy is in good agreement with our explicit supercell calculations, the deviation being
of 3\% for the smallest supercell [Fig.~2(f) and (g)].

Since Li$_2$O$_2$ exhibits localized polarons in fairly small supercells, in Fig.~\ref{fig2}(f) we compare
the wavefunction obtained within our method and that obtained via an explicit DFT calculation using the 
self-interaction correction scheme described in Ref.~\onlinecite{Sio2018}. Apart from the slight asymmetry 
in the wavefunction obtained via the DFT calculation, the two results are in very good agreement. 
Furthermore, our result is essentially identical to previous findings 
based on hybrid functional calculations~\citep{Kang2012}. The agreement with explicit DFT calculations
validates {\it a posteriori} our initial assumption leading to Eq.~(\ref{eq.1}). 
Taken together, the results in Fig.~\ref{fig1} and \ref{fig2} indicate that our theory is able to 
describe both {\it large} and {\it small} polarons on the same footing. A more in-depth analysis of the 
formalism and additional tests are provided in Ref.~\onlinecite{Sio2018}.

In summary, we developed a theoretical and computational approach that allows us to investigate, 
for the first time, polaron energies and wavefunctions across the length scales, without resorting to 
supercell calculations. Our work opens up many new directions in polaron physics. For example the spectral 
decomposition encoded in the Fourier amplitudes $A_{n\bk}$ and $B_{\bq\nu}$ could be used to construct 
model Hamiltonians with realistic materials parameters. This additional step will make it possible to 
supplement DFT with path-integrals or diagrammatic Monte Carlo calculations~\citep{Feynman1955,
Prokofev2008}, and ultimately open the way to predictive {\it ab initio} many-body calculations 
of polarons in real materials.

\begin{acknowledgments}
This work was supported by a postgraduate scholarship from the Macau SAR Government (W.H.S.),
by the Leverhulme Trust (Grant RL-2012-001), the UK Engineering and Physical Sciences Research Council 
(grants No.  EP/L015722/1 and EP/M020517/1), the Graphene Flagship (Horizon 2020 Grant No. 785219 - 
GrapheneCore2), the University of Oxford Advanced Research Computing (ARC) facility 
(http://dx.doi.org/810.5281/zenodo.22558), the PRACE-17 resources MareNostrum at BSC-CNS. 
\end{acknowledgments}

\bibliography{cite.bib}

\newpage

\newpage

\end{document}